# ATP-Independent Entropy-Driven dsRNA Unwinding by DDX3X Revealed by Coarse-Grained Simulations and Deep Learning

Kang Wang[1], Chun-Lai Ren[1,2,*], &Yu-Qiang Ma[1,2]

[1]National Laboratory of Solid State Microstructures and Department of Physics, Collaborative Innovation Center of Advanced Microstructures, Nanjing University, Nanjing 210093, China
[2]Hefei National Laboratory, Hefei 230088, China
Email: chunlair@nju.edu.cn (C.-L. R.)

## Abstract

DEAD-box RNA helicases (DDXs) are traditionally known as ATP-dependent motors that unwind double-stranded RNA (dsRNA). Recent experiments, however, show that some DDXs promote dsRNA unwinding even in the absence of ATP, raising a fundamental question about the physical mechanism underlying ATP-independent strand separation. Here, we develop a minimal, physics-based coarse-grained RNA model and incorporate weak, specific interactions between DDX3X and dsRNA, revealing the inherently stochastic nature of unwinding events. The unwinding process must overcome an energy barrier, but thermal fluctuations and entropy gain provide a driving force for RNA remodeling. We identify that dsRNA separation proceeds through rare yet obligatory strand-displacing intermediates facilitated by DDX3X. By combining deep learning-assisted analysis, we further rank the contributions of different entropic components, revealing hydrogen bonding as the dominate factor, followed by base stacking and then the backbone conformation. These findings reveal a previously unrecognized physical mechanism for RNA duplex unwinding and offer an effective framework for studying RNA remodeling kinetics.

## Introduction

DEAD-box helicases (DDXs), one of the largest and most evolutionarily conserved families of RNA helicases, are present in all kingdoms of life[1]. A representative member of this family, DDX3X, is an ATP-dependent RNA helicase that regulates multiple aspects of RNA metabolism, including transcription[2], splicing[3], transport[4], translation[5], and degradation[6]. Given that many RNA metabolic processes require the establishment of specific RNA structures, the helicase core of DDX3X—comprising its two RecA-like structural domains (D1 and D2)—unwinds short double-stranded RNA (dsRNA) into single-stranded RNA (ssRNA)[7]. Dysfunctional variants of DDX3X are etiologically linked to multiple human pathologies, most notably the neurodevelopmental disorder termed DDX3X syndrome[8]. A recent study by Owens et al[9]., demonstrated that pathogenic mutations in DDX3X compromise RNA unwinding capacity and translational efficiency, suggesting a plausible mechanism underlying these human pathologies.

The central role of DDX3X in cellular homeostasis has spurred extensive studies on its RNA remodeling mechanism. He et al.[7] showed that two DDX3Xs cooperatively unwind a 23-bp dsRNA: each domain binds ssRNA, with conformational changes triggered by ATP that allow unwinding. Toyama et al.[10] further demonstrated the stronger affinity of DDX3X for ssRNA over structured motifs, underpinning its strand separation activity. DDXs thus employ a localized strand separation mechanism that involves direct duplex binding and ATP-dependent melting, different from the processive translocation seen in DNA helicases[11]. Strikingly, Yanas et al. reported that DDX3X unwinds structured RNAs without ATP[12], raising a fundamental question: how does ATP-independent unwinding occur? This mechanistic puzzle presents substantial experimental hurdles, underscoring the need for effective coarse-grained models of RNA-protein complexes.

Current RNA models, ranging from simplified semi-flexible chain representations[13] to more detailed Martini3[14] and oxRNA[15] models, struggle to effectively simulate RNA unwinding process due to insufficient resolution or limitations in simulation timescales. To bridge this gap, we developed HelixTriad coarse-grained RNA model by adopting the three-bead-per-nucleotide framework proposed in previous studies[16, 17, 18], explicitly incorporating hydrogen bonding and base-stacking interactions. It effectively captures the conformational transitions observed experimentally during dsRNA melting, quantitatively reproduces experimental melting temperatures. It also yields persistence lengths consistent with experimental measures and reproduces the semiflexible conformations of dsRNA. Furthermore, this model can be extended to RNA-protein complexes. Using Langevin dynamics simulations, we investigated both DDX3X-dsRNA binding and unwinding dynamics. Our results reveal that dsRNA unwinding occurs stochastically and is driven by entropy gains under thermal fluctuations, a mechanism closely mirroring the canonical DNA denaturation model proposed by Jensen and Hippel in 1976[19]. Notably, successful unwinding trajectories involve crossing an energy barrier that arises from the competition between enthalpy and entropy. Given that entropy is inherently unmeasurable, we turned to machine learning to identify the key factors governing its change. We developed Entropy-Unet and demonstrated its capability to predict entropy variations during the unwinding of dsRNA under the influence of DDX3X. Furthermore, this method can dissect the distinct contributions of various interactions within the complex system to the overall entropy change, providing new insights into the fundamental nature of entropy.

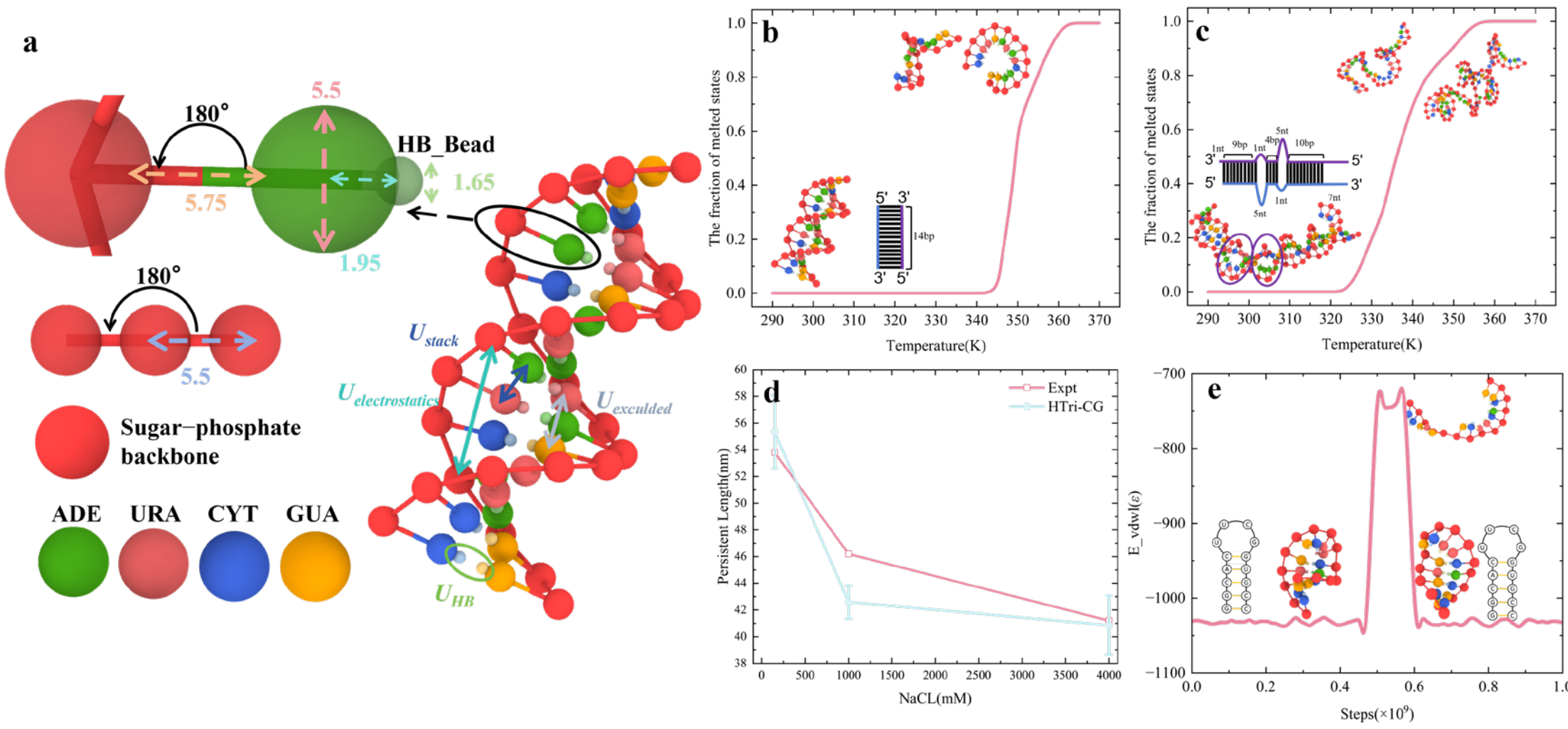

Fig 1. Development of HTri-CG RNA model and predication of dsRNA melting curves. (a) Schematic diagram of the HTri–CG RNA model architecture. Units: see Supplementary Information's Length-scale mapping. (b)–(c) Thermodynamic characterization of dsRNA denaturation (28-nt S1S2 and 73-nt XY) using REMD. The fraction of melted states is plotted as a function of temperature (290–370 K). Representative local configurations are shown for each dsRNA system. The purple circle in (c) highlights two interloop structures in the XY duplex. (d) The persistence lengths of 20-bp dsRNA from equilibrated MD trajectories at three NaCL concentrations (150 mM, 1000 mM, and 4000 mM). (e) Time-dependent evolution of vdW interaction energies at 300K, with insets illustrating three-dimensional hairpin conformations and corresponding secondary structure diagrams sampled at three distinct temporal phases.

## Results

### Development and Validation of the HTri–CG Model

To develop a model suitable for studying RNA unwinding, we built upon a three-bead-per-nucleotide framework[16, 17] and parameterized three distinct interaction sites to explicitly incorporate hydrogen bonding, base stacking and phosphodiester backbone electrostatics, yielding HelixTriad coarse-grained (HTri–CG) RNA model (Fig.1a). This model captures conformational rearrangements during RNA melting and supports long-timescale simulations. Additional methodological details can be found in the Coarse–Grained Modelling of Supplementary Information (SI).

In vitro studies have shown that DDX3X can bind and unwind 13–19 bp dsRNA[20]. To validate the HTri-CG RNA model, we characterized the melting curve and flexibility of two dsRNA duplexes (see SI's section of dsRNA and Hairpin constructs): a 14-bp (28-nt) duplex composed of strands S1 and S2 and a 23-bp (73-nt) duplex consisting of strands X and Y. The simulation box was set to $30^3$ $nm^3$ and contained ten dsRNA molecules to ensure statistical reliability. Replica exchange molecular dynamics (REMD, see SI's Replica exchange molecular dynamics Implementation) simulations were employed to enhance sampling efficiency. Melting curves were obtained by statistically analyzing simulation snapshots to distinguish between hybridized and melted (≥60% HB broken) states, as defined in Fig.1b-c. Both the 14-bp S1S2 and 23-bp XY dsRNAs exhibited three distinct regimes:(1) Stable Hybridization (290–345 K for S1S2; 290–325 K for XY): The duplexes maintained stable hybridized states. (2) Partial Melting (345–359 K for S1S2; 325–355 K for XY): A coexistence of hybridized and melted states was observed. (3) Complete Melting (≥359 K for S1S2; ≥355 K for XY): The fully melted state became dominant. The melting temperature ($T_m$), corresponding to 50% population of melted states, was determined to be 349.5 ± 0.3 K for S1S2 (experimental reference value $T_{exp}$ = 349.4 K[21]) and 336.5 ± 1.0 K for XY ($T_{exp}$ = 337.4 ± 0.5 K[22]). Given that the experimental melting curves were measured under dilute-solution conditions, we reduced the number of dsRNA molecules in the system from ten to six. The resulting melting curves remained essentially unchanged (SI Fig.S1).

Our structural analysis revealed two predominant conformations of ssRNA in the melted state: single-helical strands and open-ring structures (Fig.1b-c). The open-ring configuration exhibited remarkable persistence mediated by base-stacking interactions, which are known from our experience to predispose such structures toward hairpin formation during kinetic processes. This observation prompted validation of the semi-flexibility parameterization in our coarse-grained RNA model. We calculated the bending persistence length (P) of 20-bp dsRNA from equilibrated MD trajectories at three NaCL concentrations(150 mM, 1000 mM, and 4000 mM), following Dong et al.[23]. For each concentration condition, 300 fragments of 13-bp (Lc, θ<30°) dsRNA (excluding three terminal base pairs) were randomly sampled; bending angles were binned at 0.5°, and the distribution $p(\theta)$ was fitted to the equation $-ln(p(\theta)/sin\theta) = P\theta^2/2L_c$, yielding P values in good agreement with experiments (Fig.1d). To further validate the model for short ssRNA, we performed extensive MD simulations (total 1 × $10^9$ steps) starting from a folded 14-nt tetraloop hairpin with a 5-bp stem (Fig.1e). While the hairpin is energetically favored from a van der Waals (vdW) perspective (including base stacking, hydrogen bonding, and steric effects), consistent with Nozinovic et al.[24], it transiently unfolds into an open-ring structure and refolds within the duration of the simulation. This reversible transition confirms that the semi-flexibility parameters of our HTri-CG RNA model are well calibrated. Additionally, SI Table S1 tabulates ideal A-RNA structural data for comparison, with our HTri–CG RNA exhibiting favorable consistency. Overall, those results that are in close agreement with experimental melting temperatures, persistence lengths, and the diverse RNA conformations observed further confirm that the HTri-CG Model reliably captures RNA conformational dynamics and is suited for investigating short dsRNA unwinding.

**Intrinsic Stochasticity of ATP-Independent dsRNA Unwinding by DDX3X**

With the introduction of DDX3X, an important issue is describing the interactions between DDX3X and dsRNA. Previous work has established that their interactions are primarily mediated by specific recognition between RecA-like D1/D2 domains and RNA nucleotide phosphate groups and sugar moieties[7]. To quantify the binding specificity, we integrated a sticker-and-spacer model[25]. The binding affinity was determined using a systematic titration approach. We progressively increased the interaction strength between the D1/D2 domains and the Sugar-Phosphate Bead (SPB) in increments of $0.72\varepsilon$ (where $\varepsilon$ is the binding energy in kal/mol) until observable dissociation events occurred. This procedure yielded a measured binding energy of $10.00\varepsilon$ between dsRNA and the D1/D2 domains of DDX3X. All coarse-grained simulations were performed in the NVT ensemble using a Langevin thermostat at room temperature. Each simulation system consisted of a single DDX3X and one 14-bp dsRNA molecule placed in a cubic box with periodic boundary conditions. To mimic a dilute solution and assess finite-size effects, we used two different box sizes: $15^3nm^3$ and $30^3nm^3$. For each box size, 50 independent replicas were conducted. The simulations were preformed using the LAMMPS software package[26]. Random initial trials yielded diverse outcomes within the simulation duration, which we categorized into four distinct states based on the observed unwinding extent of the dsRNA:

(i). Stable Bound (SB): The dsRNA remained fully base-paired and stably complexed with DDX3X.

(ii). Partially Unwound (PU): DDX3X partially disrupted the dsRNA duplex, leading to localized strand separation.

(iii). Fully Unwound (FU): DDX3X completely dissociated the dsRNA into two single-stranded RNAs.

(iv). Partially Re-closed (PR): A special category of the PU state. This state formed from a FU via subsequent re-annealing of the separated strands, resulting in a partially paired structure.

The existence of four distinct conformational states implies that dsRNA unwinding is a stochastic process. To quantify the unwinding efficiency, we calculated the dwell probabilities of these final states, averaged over 50 independent replicas, shown in Fig.2a. Due to the interconversion between the FU and PR states, they were analyzed as a combined state (FU&PR). In both simulation box sizes, successful unwinding events were rare, with a low transition probability of 0.18 for the complete unwinding pathway SB→PU→FU. Among the four states, PU is the most probable. Given that dsRNA unwinding involves multiple steps of conformational change, we further calculated the probability of each transition step. These probabilities, referred to as transition rates, are presented in Fig.2b. Notably, the transition PU→FU has the lowest probability, indicating that PU is the primary kinetic bottleneck in the unwinding process. In addition, because dsRNA first binds to DDX3X before unwinding, we defined the time from the start of the simulation to the onset of unwinding as the pre-unwinding duration. In Fig.2c, the x-axis represents the pre-unwinding duration, while the y-axis indicates the number of simulations in which no conformational transition had occurred by that time. These resulting distributions exhibit Poisson-like characteristics, supporting that the onset of unwinding is stochastic.

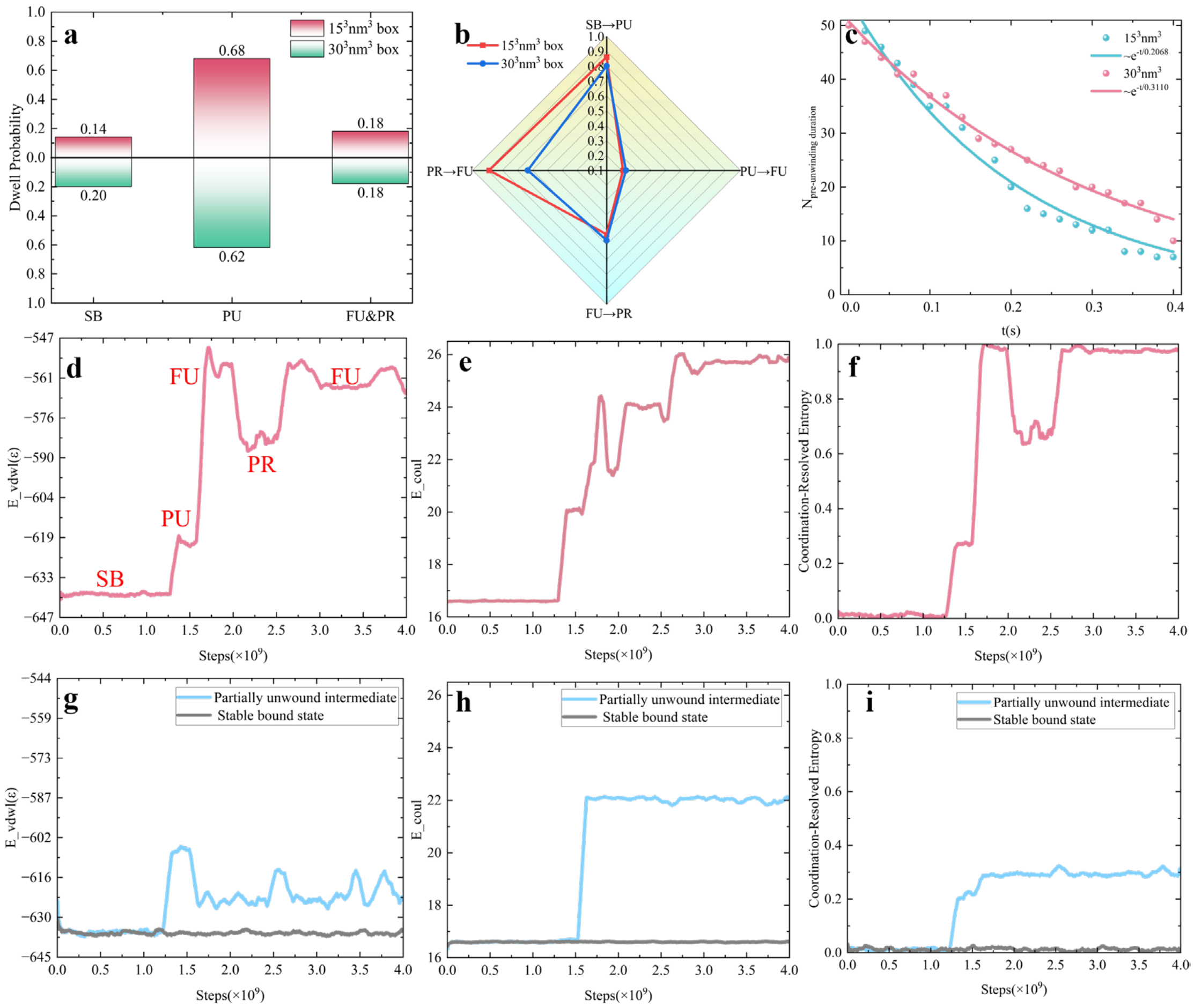

Fig 2. Stochastic, entropy-driven dsRNA unwinding mediated by DDX3X. (a) State dwell probabilities (SB, PU, FU&PR); (b) population of transition among the four transition pathways (SB→PU, PU→FU, FU→PR, PR→FU); and (c) distributions of pre-unwinding durations, averaged over 50 independent replicas for two box sizes. (d)~(f) and (g)~(i) present the analyses of vdW, Coulombic forces, and CRE for one successful unwinding event and for two abortive unwinding attempts, respectively, conducted in the $15^3$ nm$^3$ simulation system.

It's noteworthy that the SB state demonstrates RNA clamping functionality, characterized by prolonged RNA binding capacity[27]. This observation aligns with established experimental evidence demonstrating that DDXs can form stable, long-lived RNA complexes[28]. The structure of the PU conformation is investigated by Toyama et al., using $^{19}$F probe at 2023[10], identifying it as a transient intermediate during RNA unwinding. Furthermore. PU involved 4-8bp disruptions in 100 independent MDs, revealing the variable nature of this intermediate state. The PR state exhibits unstable unwinding activity of DDX3X, requiring persistent disruption of the energy barriers maintained by base stacking and hydrogen bonding interactions. Specifically, this destabilization predominantly occurs at terminal base pairs (2-3bp), revealing position-dependent stability patterns during duplex reformation.

## Entropy-Driven dsRNA Unwinding by DDX3X

To investigate the primary driving mechanism behind the successful unwinding of dsRNA, we examined the system's enthalpic and entropic contributions to elucidate their respective roles. Coordination-resolved entropy (CRE; see Materials and Methods) effectively captures the conformational changes accompanying double-strand separation. Fig.2d-f depict the temporal evolution of vdW interactions, Coulombic potentials, and CRE changes during a successful unwinding event, which proceeds along a specific pathway: SB → PU → FU → PR → FU. As the bases separate, vdw attractions weaken, whereas preferential binding of DDX3X to the ssRNA backbone brings RNA backbone beads into closer proximity, thereby strengthening Coulombic interactions. The resultant net positive enthalpy change ($\Delta H>0$) thermodynamically disfavors spontaneous unwinding. However, progressive entropy gains ($\Delta S>0$) across successive states (SB→PU→FU→PR→FU) overcome the enthalpic barrier, confirming the process as entropy-driven. The entropy compensation mechanism originates from increased conformational freedom in ssRNA strands.

Although successful unwinding can occur, unsuccessful cases still abundantly exist: transitions from the SB to PU states that result in persistent PU state, as well as prolonged stability in the SB state. Fig.2g-i illustrates these dynamics, with sky blue curves representing PU state transitions and gray curves denoting SB state behavior. Notably, the PU state (sky blue) demonstrates a characteristic vdW interaction surge, beginning around $\sim 1.25\times 10^9$ steps. Structural analysis of snapshots reveals concurrent disruption of 3-bp during this stage. Subsequently, the number of disrupted base pairs fluctuates before stabilizing, leaving only eight base pairs intact in the PU state. Coulomb interactions exhibited minimal fluctuation, following a step-like behavior corresponding to the sequential opening of base pairs. CRE effectively tracked this progressive process with high sensitivity. In contrast, the SB state (gray) maintains stable across all parameters (vdW, coulomb interactions, and entropy), keeping the RNA clamping throughout the process—a futile unwinding attempt during the unwinding cycle.

**Unwinding Dynamics and Energy Barrier of DDX3X-dsRNA Complex**

To explore the unwinding dynamics, we employed *Distance* and *Rg*, two key metrics derived from MD trajectories. *Distance* represents the center-of-mass distance between the two strands of dsRNA, while *Rg* quantifies the overall compactness of the DDX3X-dsRNA complex. A representative successful unwinding trajectory was chosen to illustrate the conformational dynamics during the process. Fig. 3a-b show the evolutions of *Distance* and *Rg*, with several representative states (I-VII) highlighted along the trajectory. For clarity in visualizing structural transitions, the intrinsically disordered N-terminal region (IDR) of DDX3X was omitted in Fig.3c.I-VII, as it was not associated with the dsRNA during the unwinding process (SI Fig.S2). Consistent with our observations, previous studies have also reported minimal IDR binding to poly-$U_{10}$ ssRNA, GC-14mer dsRNA, and a 14mer tetraloop RNA[29]. The system follows a sequence of conformational transitions: I → II → III → IV → V → VI → VII.

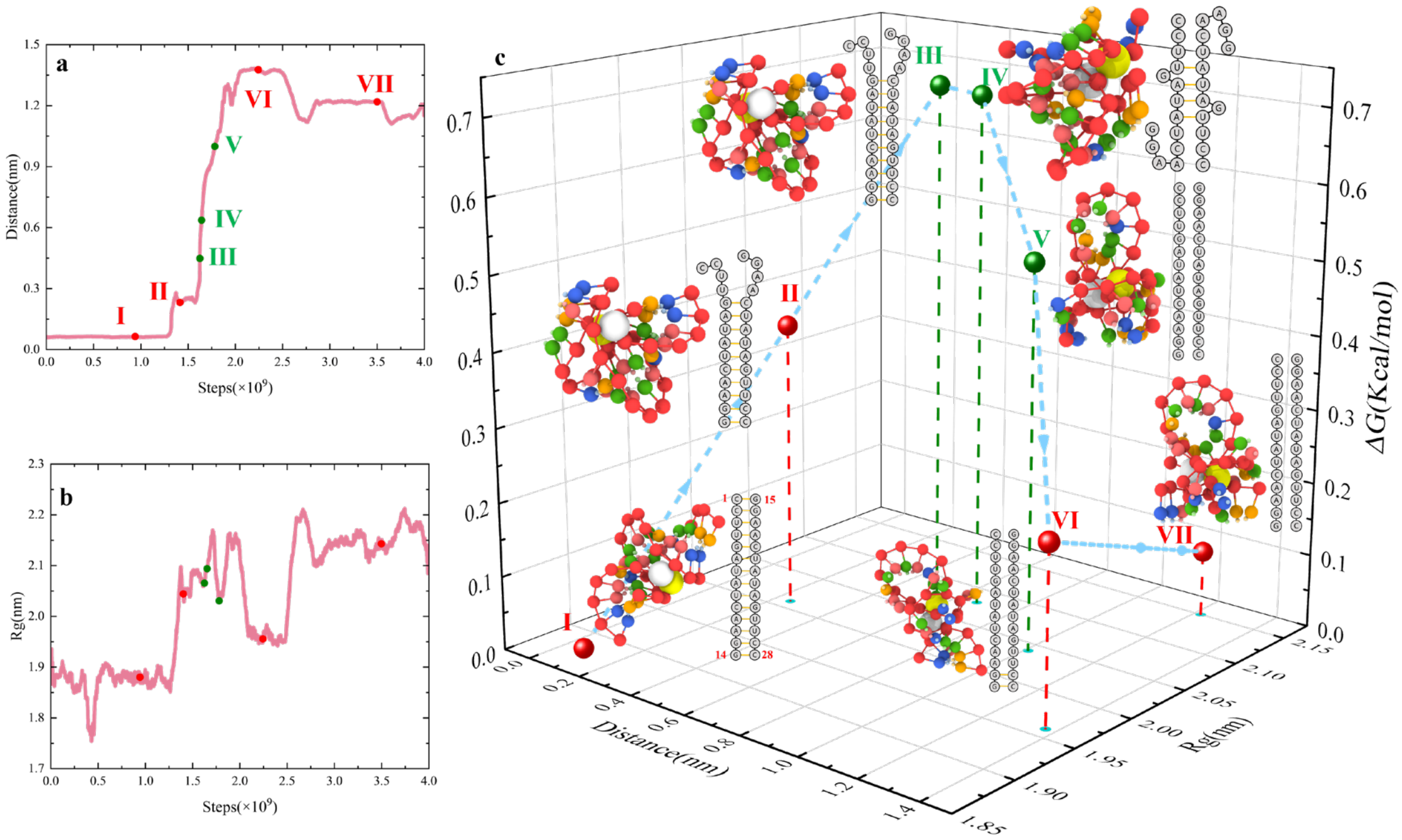

Fig 3. Dynamics of unwinding dsRNA by DDX3X. (a) Temporal evolution of Rg. (b) Temporal evolution of Distance. Sphere color encodes the structural stability of the DDX3X-dsRNA complex: red indicates low-energy, long-lived states, whereas green corresponds to higher-energy, transient states with short-lived lifetime. (c) Representative conformational snapshots along the unwinding trajectory at selected time points. I: $9.39\times10^8$ step; II: $1.4\times10^9$ step; III: $1.625\times10^9$ step; IV:$1.6432\times10^9$ step; V: $1.7824\times10^9$ step; VI: $2.2441\times10^9$ step; VII: $3.5\times10^9$ step. The visualizations integrate three-dimensional complex topology (with the disordered regions of DDX3X omitted for clarity).

In SB state shown in Fig.3c.I, structural visualization of the DDX3X-dsRNA complex demonstrated tight binding of DDX3X to dsRNA, inducing localized bending, and secondary structure of dsRNA remains intact. State Fig.3c.II corresponds to the PU state, characterized by the disruption of 4-bp and bent ssRNA segments that preferentially interact with the D1 and D2 helicase domains, thus exposing the base of the groove regions. The transient state in Fig.3c.III also represents the PU state, but exhibits complex base rearrangement dynamics mediated by DDX3X under thermal fluctuations. In particular, transient disruptions of hydrogen bonding networks and base stacking cause multiple base pair destabilizations, visualized as an unstable configuration. Concomitantly, bending of ssRNA backbone facilitates helicase binding, while the dsRNA adopts hybrid conformations that incorporate both helix and interloop elements.

Progression to the unstable state in Fig.3c.IV was marked by sequential disruption of 3-bp in terminal, with ssRNA adopting circular conformations around D1/D2 domains. This structural rearrangement preserved the four central base pairs in helical configuration, maintaining only short helix elements in secondary structure. The FU state in Fig.3c.V showed complete helical disintegration, with circularized ssRNA strands encapsulating D1/D2 domains and fully exposed bases. The rapid III→IV→V evolutions correlated with abrupt increases in Distance during unwinding (Fig.3a), indicating a successful unwinding event. Post-unwinding relaxation enabled structural reorganization into FU conformation in Fig.3c.VII. Subsequent trajectory revealed reformation of PR state in Fig.3c.VI through terminal RNA base stacking and hydrogen bonding, followed by reversion to FU state in Fig.3c.VII. The interconversion between FU and PR states is maintained by continuous enthalpy-entropy compensation arising from thermal fluctuation. Successful attainment of the FU state requires passing through the transient intermediates (Fig.3c.III and IV), which correspond to unstable, high-energy strand-displacement states.

By comparing successful and unsuccessful unwinding events, we found that successful transition from the SB state to the PU state requires capturing stochastic nucleobase fluctuations at either dsRNA terminus, a process that is independent of DDX3X binding sites. The transition from the PU state to the FU state can proceed via two distinct pathways (SI Fig.S3): (i) In the SB state, sequential opening of 3-4 base pairs forms the PU state (Fig.3c.II), followed by continuous base pair disruption (maintaining 5-8 opened pairs). (ii) The PU state (Fig.3c.II) undergoes base-rearrangement to form another PU state (Fig.3c.III) and induces rupture of terminal base pairs (Fig.3c.IV), ultimately resulting in complete dsRNA unwinding. This bifurcation mechanism underlies the stochastic nature of DDX3X-mediated dsRNA unwinding, where pathway selection depends on real-time capturing of RNA nucleobase fluctuations.

Notably, the free energy of different states ($\Delta G$; see Materials and Methods) is calculated, and the energy barrier is constructed in a two-dimensional reaction coordinates space defined by *Distance* and $Rg$ (Fig. 3c). This behavior indicates that the underlying physical mechanism of the unwinding process shares key features with a nucleation-like transition. Successful unwinding corresponds to trajectories that cross this free-energy barrier, whereas trajectories that fail to overcome the barrier instead relax into PU or SB states. This two-dimensional representation highlights that unwinding is not governed by a single structural variable, but instead arises from the coupled evolution of strand separation and global conformational rearrangement. These results implies that thermal fluctuations and entropy gain can provide a driving force for dsRNA remodeling, enabling stochastic barrier crossing.

**Development and Evaluation of Entropy-UNet**

Artificial Intelligence (AI) has emerged as an indispensable analytical tool across diverse scientific fields, facilitating cross-disciplinary integration through advanced data-processing capabilities. In non-equilibrium dynamic processes, information entropy plays a fundamental role in determining the direction of evolution, yet it is often difficult to quantify. In this context, coordination-resolved entropy (CRE) provides a means to capture the conformational changes associated with dsRNA unwinding driven by DDX3X. Motivated by this, we aimed to develop a deep learning framework to predict CRE and to dissect the distinct contributions underlying its behavior.

Entropy-UNet (Fig.4a,b) was developed to leverage coarse-grained trajectories generated by Metadynamics[30] (see SI's Metadynamics Implementation) for constructing physically informed entropy fingerprints. This approach enables precise mapping between transient conformational states and entropy variations. Metadynamics served as the primary sampling strategy, employing predefined CVs to effectively capture conformational states for deep learning applications. We generated two distinct datasets using $Rg^{system}$ and Dist as CVs, yielding 40,000 and 30,000 conformational samples, respectively. These datasets were systematically partitioned into training, validation, and test subsets at an 8:1:1 ratio. The model underwent 200 training epochs with a learning rate of 0.001, demonstrating robust performance on test data (KLD=0.0887; CE=3.8189; BS=0.0014; CS=0.9727).

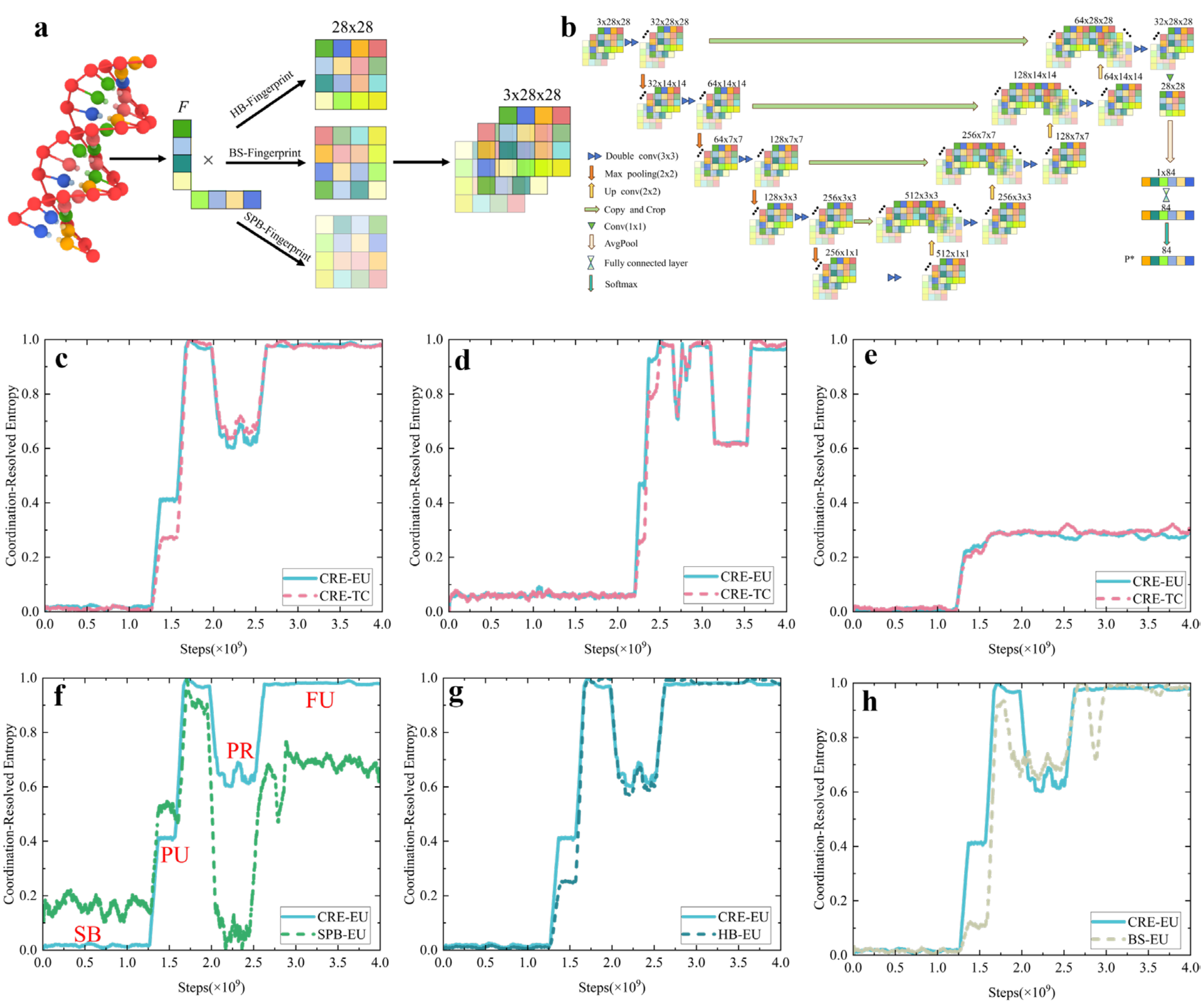

Fig 4. Development and evaluation of Entropy-UNet. (a) Schematic of entropy fingerprints. (b) Entropy-Unet Architecture Overview. (c)~(e) Comparative Analysis of CRE-EU and CRE-TC across three independent test datasets. (f)~(g) Entropic contributions of SPB, HB, and BS were Entropy-Unet resolved.

For rigorous validation, we evaluated three independent test cases. Two diverse successful unwinding trajectories (Fig.4c,d) with significant conformational transitions, and an unsuccessful trajectory (Fig.4e) persistently occupying the PU state. Quantitative comparison between CRE-EU (CRE predicted from Entropy-Unet) with CRE-TC (CRE-Theoretical Calculations) curve revealed strong predictive accuracy. Fig.4c illustrates the complete SB→PU→FU→PR→FU transition pathway during dsRNA unwinding by DDX3X. Entropy-Unet precisely tracked entropy changes across states, with minor overestimations observed for PU states in Fig.4c and d. Notably, predictions for PU states in test cases Fig.4e showed excellent agreement with theoretical CRE values.

To evaluate whether Entropy-UNet captures meaningful physical interactions, we analyzed its architecture to assess how Hydrogen Bonding (HB), Base Stacking (BS), and Sugar-Phosphate Bead (SPB) fingerprints influence PU state predictions. The encoder's final bottleneck layer compresses spatial fingerprints into high-level semantic representations, while the decoder up-samples with residual skip-connections. Guided by this, we applied Grad-CAM[31] to the penultimate encoder layer to examine entropy fingerprint activation. Three representative PU state instances were analyzed (SI Fig.S4), revealing that SPB features exhibit higher semantic sparsity than HB and BS. Incomplete SPB representations correlated with prediction errors, causing over- or underestimation of entropy in PU and PR states (Fig.4c). These findings indicate that Entropy-UNet primarily learns physically meaningful interactions, including BS, HB, and backbone electrostatics, while effectively filtering non-essential background features.

Overall, Entropy-UNet demonstrated consistent predictive reliability across all test conditions, establishing a novel framework linking conformation with entropy. Notably, in Fig.4d, the model captured complex transition dynamics, including extended SB occupancy followed by rapid PU→FU→PR→FU conversion, highlighting its structural resolution capability. Using "soft pixel" feature images, we compared Entropy-UNet with a baseline CNN (without residual connections) and a Transformer-Encoder[32] on an independent test set (SI Fig.S5). While overall performance was comparable to the CNN, Entropy-UNet excels at capturing biophysically meaningful interactions, extending advances in structure-aware deep learning exemplified by UFold[33]. Although the Transformer-Encoder better identified PU states, it underperformed for PR states. In contrast, Entropy-UNet maintains robust, balanced performance across conformational states, providing consistent interpretability and reliability for structural bioinformatics applications.

**Distinct contributions of CRE predicted by Entropy-Unet**

The CRE metric employs coordination numbers of dsRNA beads as spatially resolved indicators of local structural order, quantifying entropy through cooperative interactions among proximal particles. This approach inherently captures the coupled thermodynamic contributions of hydrogen bonding, base stacking, and sugar-phosphate backbone electrostatic interactions. Entropy-Unet employs entropy fingerprints explicit spatial adjacency modeling between bead types to dynamically resolve their distinct spatial information. To disentangle these effects, we systematically evaluated their individual contributions to CRE by computing independent CRE values for each component. Our analysis reveals three key findings (Fig.4f-h): First, SPB-EU successfully distinguished SB and PU states, albeit with systematic overestimation, while detecting PR and FU states with consistent underestimation based on entropy profile analysis; Second, HB-EU demonstrated exceptional agreement with reference values, accurately identifying all four states (SB, PU, PR, FU) with only minor PR-state underestimation; Third, BS-EU showed intermediate performance, resolving all states with moderate accuracy (superior to SPB-EU but inferior to HB-EU) while slightly underestimating PU-state entropy. Leveraging Entropy-Unet's precise, interaction-specific entropy decomposition, we established the thermodynamic hierarchy of contributions: hydrogen bonding (primary) > base stacking (secondary) > backbone electrostatics (tertiary), with cooperative HB and BS interactions dominating the overall entropy landscape.

## Discussion and conclusion

In this study, our integrated computational approach, combining the coarse-grained HTri-CG RNA model with Entropy-Unet, reveals that DDX3X-mediated dsRNA unwinding in the absence of ATP occurs via a stochastic, entropy-driven process. The HTri-CG RNA model reveals that the unwinding dynamics is primarily governed by DDX3X-induced disruption of Watson-Crick base pairing and base stacking interactions through thermal fluctuations, which facilitates relative motion between the two ssRNA strands and promotes structural melting by overcoming high-energy intermediate states. Entropy-Unet's high-resolution decomposition further shows that hydrogen bonding and base stacking interactions are the primary determinants affecting RNA conformational reorganization during DDX3X-mediated dsRNA unwinding. Remarkably, this mechanism mirrors well-characterized DNA melting processes, such as those mediated by bacteriophage T4 gene 32-protein[34], highlighting its generalizability to RNA systems. Meanwhile, this unwinding mechanism aligns with the theoretical framework of passive unwinding of helicases proposed by Thirumalai et al[35, 36]. It further reveals a significant evolutionary strategy: for helicases lacking powerful active disruption capabilities, utilizing environmental thermal fluctuations for unwinding represents an efficient and conserved physically economical solution.

While our integrated approach successfully reproduced experimental observations of DDX3X-mediated dsRNA unwinding, several methodological disadvantages and advantages should be acknowledged:

(i). HTri-CG RNA Model:
Compared to the Martini3 and oxRNA models, the HTri-CG model sacrifices some structural precision, including the accurate description of base stacking and hydrogen bond anisotropy. Nevertheless, its reliability has been validated across multiple RNA systems, including a short helical S1S2(28nt) dsRNA, a long helical XY (73nt) dsRNA with two interloop elements, and a 5-bp stem tetraloop hairpin. Although the model does not explicitly represent non-canonical base pairs, this omission is justified empirically. Taken together, the HTri-CG model offers an effective balance between accuracy and efficiency for simulating RNA unwinding dynamics.

(ii). Modeling Protein-RNA Interactions:
According to He et al.[7], interactions between the D1/D2 domains and nucleic acids are the primary determinants of DDX3X-mediated unwinding. This insight enabled the successful simulation of dsRNA unwinding by DDX3X. The coarse-grained model of DDX3X was constructed using the sticker-and-spacer framework (SI Fig.S6), a simplified representation known for its scalability. When extending this DDX3X-RNA framework to other protein–RNA complexes—such as the helicases DDX5[37] and eIF4A[38]—precise definition of the interaction sites and binding modes between the helicase and nucleic acid becomes essential. Consequently, the approach depends heavily on the availability of high-resolution crystal structures of such complexes. Recent advances in predictive accuracy for RNA–protein interactions, achieved through algorithms like AlphaFold3[39], may help mitigate this limitation.

(iii). Entropy-UNet Performance:
Entropy-UNet utilizes HB, BS, and SPB entropy fingerprints to effectively characterize the three primary RNA interaction types, enabling both accurate entropy prediction and entropy contributions decoupling (Hydrogen bonding > Base stacking > Backbone electrostatics). However, while the architecture maintains translation invariance and hierarchical feature learning capabilities, it lacks explicit rotational symmetry, potentially limiting its robustness when analyzing PU conformational ensembles. Furthermore, the dimensionality of entropy fingerprints is inherently constrained by the bead count in the coarse-grained dsRNA representation.

In our future work, we will continue to delve into RNA-protein complex systems, and will persistently expand our models and strategies to investigate open questions, such as RNA systems extensively featuring non-canonical structures or more complex protein condensate systems.

## Materials and Methods

### Free energy and coordination-resolved entropy

To characterize the free energy of DDX3X-dsRNA unwinding in MD simulations, we performed binning-based statistical analysis using two essential geometric parameters extracted from the MD trajectories: (1) the radius of gyration of the entire system ($Rg = \sqrt{\frac{1}{N}\sum_{i=1}^{N}|r_i - R_{com}|^2}$, $R_{com}$ is center of mass coordinate vector) and (2) the center-of-mass distance between two strands of the dsRNA ($Distance = |R_{com}^{X} - R_{com}^{Y}|$). This Visited States Method[40] directly utilized the time-evolution of these MD-calculated structural metrics to construct the free energy. While this method for obtaining complete free energy generally requires exhaustive sampling of high-barrier regions, our extended simulations ($4\times10^9$ steps) successfully captured key metastable intermediates in DDX3X-dsRNA unwinding, ensuring robust free energy reconstruction. *Rg* and *Distance* dimensions were discretized into 100 equally spaced bins per axis, generating a 10,000-bin grids. The discrete joint probability distribution $P\,(Rg_i, Distance_j)$ for bin ($i,j$) was calculated as:

$$P\left(Rg_i, Distance_j\right) = \frac{N_{ij}}{N_{total}} \quad (1)$$

where $N_{ij}$ denotes the number of MD frames falling into the bin $(i, j)$, and $N_{total}$= 4.0×10$^4$ corresponds to the total number of frames sampled from MD trajectories. Free energy was subsequently derived via:

$$\Delta G = -0.001 N_A K_B T ln \frac{P\left(Rg_i, Distance_j\right)}{P(Rg, Distance)_{max}} \tag{2}$$

with $T$=300K, $N_A$ is the Avogadro constant and $P\ (Rg,\ Distance)_{max}$ as the maximum probability bin to ensure that the lowest free energy $\Delta G = 0$.

Our analysis reveals that DDX3X-mediated dsRNA unwinding originates from thermal fluctuations, disrupts the hydrogen-bonding network and base stacking interactions, then facilitating structural melting through preferential binding to the single-stranded denatured. Meanwhile, the conformational changes during chain dissociation processes occur at ultrafast timescales, the subtle structural variations in these conformations remain traceable and inherently contain valuable information-particularly through coordination number alterations-that reflects thermal fluctuations. To quantify these conformational changes, we employ an information entropy framework through the following protocol:

(i) Calculate single-bead a mollified version of radial distribution functions by[41]

$$g_m^i(r) = \frac{1}{4\pi\rho r^2} \sum_j \frac{1}{\sqrt{2\pi\sigma^2}} e^{-\frac{\left(r-r_{ij}\right)^2}{(2\sigma^2)}} \tag{3}$$

where $r_{ij}$ is the distance between beads $i$ and $j$, $j$ are the neighbors of bead $i$, and σ is a broadening parameter.
(ii) Determine single-bead coordination numbers via integration of Eq. 4;

$$n_i(r) = 4\pi\rho \int_0^r g_m^i(r)\, r^2 dr \tag{4}$$

Where $\rho$ is localized density of bead, $r$=2.1nm is cutoff distance.
(iii) Establish coordination number probability of single-bead through Eq.5;

$$P(i) = \frac{n_i(r)}{\sum_i^n n_i\ (r)} \tag{5}$$

(iv)We use Shannon entropy (Eq.6) followed by normalization to obtain coordination-resolved entropy (CRE).

$$CRE = -\sum_i^n P\ (i) \log_2 P(i) \tag{6}$$

**Entropy Fingerprint**

In supervised learning, the dual challenge of feature engineering and label engineering persists, where intrinsic coupling between feature-label pairs necessitates coordinated design strategies. To address this, we developed an entropy fingerprint encoding scheme termed "soft pixels" to quantify local order of beads[41], while implementing a "soft labeling" system based on bead coordination number probability distributions in dsRNA. Entropy-UNet architecture employs a deep neural network to perform prediction of coordination-resolved entropy from input entropy fingerprints, establishing an entropic mapping between local order landscapes and coordination patterns of beads. The entropy fingerprint is obtained by

$$F_i(r) = -\sum_{i}^{n} 2\pi\rho k_B \int_0^{r_m} \left[g_m^i(r)\ln g_m^i(r) - g_m^i(r) + 1\right] r^2 dr \quad (7)$$

Where $\rho$ is localized density of system, r=3.5 nm is cutoff distance, and $k_B$ is a Boltzmann constant. Fig.4a demonstrates the computational pipeline for constructing entropy fingerprint representations. Specifically, the entropy fingerprint F of HB, BS, and SPB are initially encoded as one-dimensional vectors (length=28). Each vector undergoes tensor product transformation through Kronecker self-multiplication $F \otimes F^T$, generating 28×28 matrices that capture second-order entropy correlations, then matrices is normalized to [-1,1]. These matrices are subsequently stacked along the channel dimension to form a three-dimensional (3, 28, 28) tensor representation, termed "soft pixel images", where each channel corresponds to distinct bead-type entropy landscapes. Entropy-UNet architecture's novel input representation demonstrates three principal advantages for coordination-resolved entropy prediction:

(i). Spatially resolved interaction modeling: The tensor-based "soft pixel" representation explicitly encodes three fundamental interaction types - base stacking, base pairing, and electrostatic interactions - through distinct feature channels.

(ii). Cross-scale correlation emergence: The Kronecker product operation $F \otimes F^T$, inherently induces non-local interaction patterns from local entropy fingerprints. This mathematical transformation converts 1D bead-specific entropy vectors into 2D pairwise correlation matrices, where the off-diagonal elements *(i, j)* explicitly quantify entropic coupling between spatially separated beads *i* and *j*, thereby establishing spontaneous long-range connectivity beyond the original local descriptors.

(iii). Architectural compatibility: The grid-structured input enables implementation of fully convolutional neural networks, providing inherent advantages in processing high-dimensional conformational data through translational equivariance and hierarchical feature learning.

Other methods are described in the SI, including The Coarse-Grained Modelling, Replica exchange molecular dynamics Implementation, Metadynamics Implementation, dsRNA and Hairpin constructs and Entropy-UNet network architecture.

**Data Availability**

The data that support the findings of this study are available from the corresponding author upon request.

## Code availability

The codebase for this study are available on GitHub: https://github.com/BeastyWK/Entropy-UNet.

## Graphical TOC

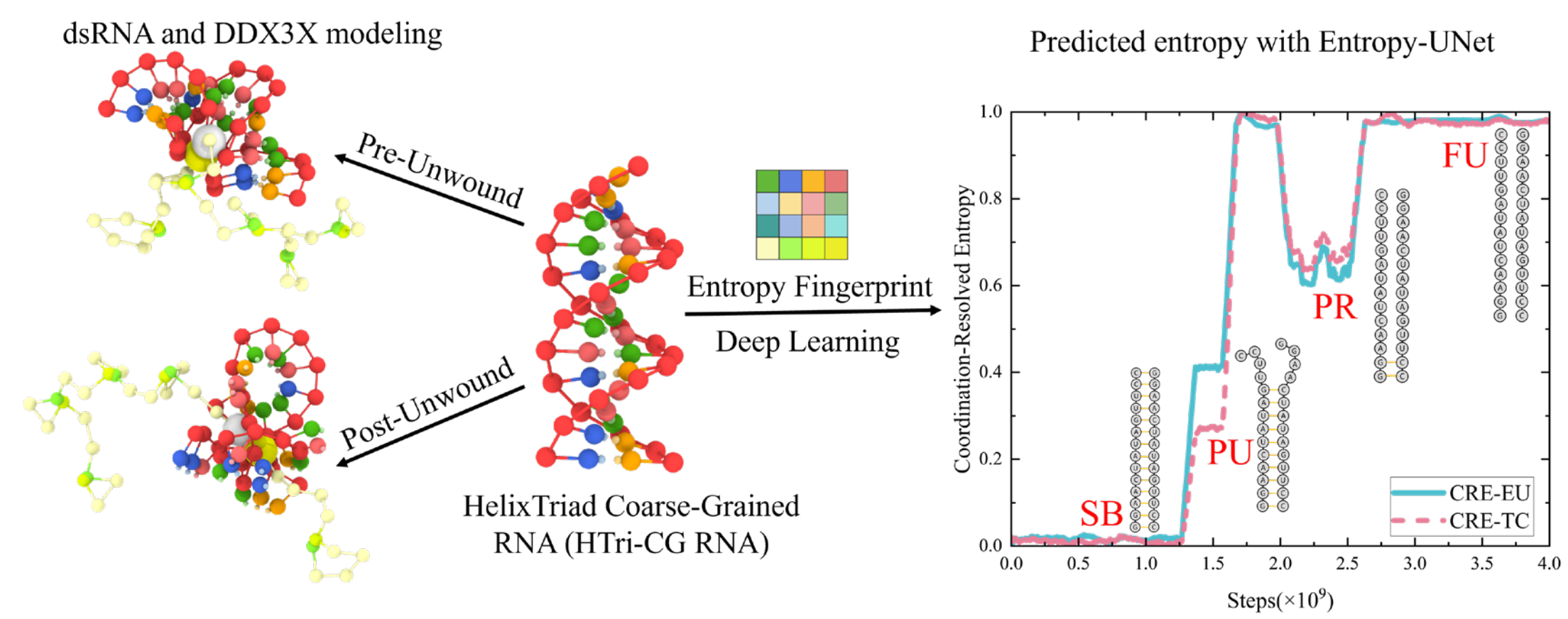

## References

1. Linder P, Jankowsky E. From unwinding to clamping - the DEAD box RNA helicase family. *Nature reviews Molecular cell biology* **12**, 505-516 (2011).
2. Chao CH, Chen CM, Cheng PL, Shih JW, Tsou AP, Lee YH. DDX3, a DEAD box RNA helicase with tumor growth-suppressive property and transcriptional regulation activity of the p21waf1/cip1 promoter, is a candidate tumor suppressor. *Cancer research* **66**, 6579-6588 (2006).
3. Calviello L, *et al.* DDX3 depletion represses translation of mRNAs with complex 5' UTRs. *Nucleic acids research* **49**, 5336-5350 (2021).
4. Elvira G, *et al.* Characterization of an RNA granule from developing brain. *Molecular & cellular proteomics : MCP* **5**, 635-651 (2006).
5. Oh S, *et al.* Medulloblastoma-associated DDX3 variant selectively alters the translational response to stress. *Oncotarget* **7**, 28169-28182 (2016).
6. Nussbacher JK, Yeo GW. Systematic Discovery of RNA Binding Proteins that Regulate MicroRNA Levels. *Molecular cell* **69**, 1005-1016.e1007 (2018).
7. Song H, Ji X. The mechanism of RNA duplex recognition and unwinding by DEAD-box helicase DDX3X. *Nature communications* **10**, 3085 (2019).
8. Gadek M, Sherr EH, Floor SN. The variant landscape and function of DDX3X in cancer and neurodevelopmental disorders. *Trends in molecular medicine* **29**, 726-739 (2023).
9. Owens MC, *et al.* Specific catalytically impaired DDX3X mutants form sexually dimorphic hollow condensates. *Nature communications* **15**, 9553 (2024).
10. Toyama Y, Shimada I. NMR characterization of RNA binding property of the DEAD-box RNA helicase DDX3X and its implications for helicase activity. *Nature communications* **15**, 3303 (2024).
11. Pyle AM. Translocation and unwinding mechanisms of RNA and DNA helicases. *Annual review of biophysics* **37**, 317-336 (2008).

12. Yanas A, Shweta H, Owens MC, Liu KF, Goldman YE. RNA helicases DDX3X and DDX3Y form nanometer-scale RNA-protein clusters that support catalytic activity. *Current biology : CB* **34**, 5714-5727.e5716 (2024).
13. Seol Y, Skinner GM, Visscher K. Elastic properties of a single-stranded charged homopolymeric ribonucleotide. *Physical review letters* **93 11**, 118102 (2004).
14. Yangaliev D, Ozkan SB. Coarse-grained RNA model for the Martini 3 force field. *Biophysical journal*, (2025).
15. Šulc P, Romano F, Ouldridge TE, Doye JP, Louis AA. A nucleotide-level coarse-grained model of RNA. *The Journal of chemical physics* **140**, 235102 (2014).
16. Kapoor U, Kim YC, Mittal J. Coarse-Grained Models to Study Protein-DNA Interactions and Liquid-Liquid Phase Separation. *Journal of chemical theory and computation* **20**, 1717-1731 (2024).
17. Jin L, Shi YZ, Feng CJ, Tan YL, Tan ZJ. Modeling Structure, Stability, and Flexibility of Double-Stranded RNAs in Salt Solutions. *Biophysical journal* **115**, 1403-1416 (2018).
18. Thirumalai D, Hori N, Nguyen HT. Minimal models for RNA simulations. *Current opinion in structural biology* **93**, 103107 (2025).
19. Jensen DE, von Hippel PH. DNA "melting" proteins. I. Effects of bovine pancreatic ribonuclease binding on the conformation and stability of DNA. *The Journal of biological chemistry* **251**, 7198-7214 (1976).
20. Putnam CD, Hammel M, Hura GL, Tainer JA. X-ray solution scattering (SAXS) combined with crystallography and computation: defining accurate macromolecular structures, conformations and assemblies in solution. *Quarterly reviews of biophysics* **40**, 191-285 (2007).
21. Serra MJ, Baird JD, Dale T, Fey BL, Retatagos K, Westhof E. Effects of magnesium ions on the stabilization of RNA oligomers of defined structures. *RNA (New York, NY)* **8**, 307-323 (2002).
22. Wu C, Shan Y, Wang S, Liu F. Dynamically probing ATP-dependent RNA helicase A-assisted RNA structure conversion using single molecule fluorescence resonance energy transfer. *Protein science : a publication of the Protein Society* **30**, 1157-1168 (2021).
23. Dong H-L, Zhang C, Dai L, Zhang Y, Zhang X-H, Tan Z-J. The origin of different bending stiffness between double-stranded RNA and DNA revealed by magnetic tweezers and simulations. *Nucleic acids research* **52**, 2519-2529 (2024).
24. Nozinovic S, Fürtig B, Jonker HR, Richter C, Schwalbe H. High-resolution NMR structure of an RNA model system: the 14-mer cUUCGg tetraloop hairpin RNA. *Nucleic acids research* **38**, 683-694 (2010).
25. Ren CL, Shan Y, Zhang P, Ding HM, Ma YQ. Uncovering the molecular mechanism for dual effect of ATP on phase separation in FUS solution. *Science advances* **8**, eabo7885 (2022).
26. Lammps. https://github.com/lammps/lammps/tree/stable_29Aug2024_update2.
27. Ballut L, Marchadier B, Baguet A, Tomasetto C, Séraphin B, Le Hir H. The exon junction core complex is locked onto RNA by inhibition of eIF4AIII ATPase activity. *Nature structural & molecular biology* **12**, 861-869 (2005).
28. Nielsen KH*, et al.* Mechanism of ATP turnover inhibition in the EJC. *RNA (New York, NY)* **15**, 67-75 (2009).
29. Toyama Y, Takeuchi K, Shimada I. Regulatory role of the N-terminal intrinsically

disordered region of the DEAD-box RNA helicase DDX3X in selective RNA recognition. *Nature communications* **16**, 7762 (2025).

30. Laio A, Gervasio FL. Metadynamics: a method to simulate rare events and reconstruct the free energy in biophysics, chemistry and material science. *Reports on Progress in Physics* **71**, 126601 (2008).
31. Dovletov G, Pham DD, Lorcks S, Pauli J, Gratz M, Quick HH. Grad-CAM Guided U-Net for MRI-based Pseudo-CT Synthesis. *Annual International Conference of the IEEE Engineering in Medicine and Biology Society IEEE Engineering in Medicine and Biology Society Annual International Conference* **2022**, 2071-2075 (2022).
32. Dosovitskiy A*, et al.* An Image is Worth 16x16 Words: Transformers for Image Recognition at Scale. *ArXiv* **abs/2010.11929**, (2020).
33. Fu L, Cao Y, Wu J, Peng Q, Nie Q, Xie X. UFold: fast and accurate RNA secondary structure prediction with deep learning. *Nucleic acids research* **50**, e14 (2022).
34. Jensen DE, Kelly RC, von Hippel PH. DNA "melting" proteins. II. Effects of bacteriophage T4 gene 32-protein binding on the conformation and stability of nucleic acid structures. *The Journal of biological chemistry* **251**, 7215-7228 (1976).
35. Pincus DL, Chakrabarti S, Thirumalai D. Helicase processivity and not the unwinding velocity exhibits universal increase with force. *Biophysical journal* **109**, 220-230 (2015).
36. Chakrabarti S, Jarzynski C, Thirumalai D. Processivity, Velocity, and Universal Characteristics of Nucleic Acid Unwinding by Helicases. *Biophysical journal* **117**, 867-879 (2019).
37. Barra G*, et al.* Human Helicase DDX5 is Hijacked by SARS-CoV-2 Nsp13 Helicase to Enhance RNA Unwinding. *ACS omega* **10**, 34941-34950 (2025).
38. Linder P, Fuller-Pace FV. Looking back on the birth of DEAD-box RNA helicases. *Biochimica et biophysica acta* **1829**, 750-755 (2013).
39. Abramson J*, et al.* Accurate structure prediction of biomolecular interactions with AlphaFold 3. *Nature* **630**, 493-500 (2024).
40. Christ CD, Mark AE, van Gunsteren WF. Basic ingredients of free energy calculations: a review. *Journal of computational chemistry* **31**, 1569-1582 (2010).
41. Piaggi PM, Parrinello M. Entropy based fingerprint for local crystalline order. *The Journal of chemical physics* **147**, 114112 (2017).

## Acknowledgments

We are grateful to the High-Performance Computing Center (HPCC) of Nanjing University for the numerical calculations in this paper on its blade cluster system.

## Funding

This work is supported by the National Key Research and Development Program of China (Grant No. 2022YFA1405000), the National Natural Science Foundation of China (Grant Nos. 12274212, 12347102), Quantum Science and Technology-National Science and Technology Major Project (Grant No. 2024ZD0300101), and Fundamental and Interdisciplinary Disciplines Breakthrough Plan of the Ministry of Education of Chain (Grant No. JYB2025XDXM502).

## Author contributions

K.W. performed the Software development, Methodology, Formal analysis, and Data curation. C.-L. R. and Y.-Q. M. designed and supervised the research. All the authors wrote the manuscript.

## Competing interests

The authors declare no competing interests.

## Additional information

### Supplementary Materials

**This PDF file includes:**

The Coarse-Grained Modelling
Replica exchange molecular dynamics Implementation
Metadynamics Implementation
dsRNA and Hairpin constructs
Length-scale mapping
Entropy-Unet network architecture
Figure S1 to S6
Table S1 to S5